\documentclass[12pt,a4paper]{iopart}
\usepackage[utf8]{inputenc}
\usepackage[english]{babel}
\usepackage{amssymb, amsfonts, amsthm, esint}
\usepackage{graphicx}
\usepackage{slashed}
\usepackage{color}
\usepackage[usenames,dvipsnames,svgnames]{xcolor}
\usepackage[colorlinks=true, linktocpage=true]{hyperref}
\usepackage{tabularx}
\usepackage{enumitem}
\usepackage[activate={true,nocompatibility}, final, tracking=true, kerning=true, spacing=true, factor=1100, stretch=10, shrink=10]{microtype}
\usepackage{booktabs}
\usepackage{listings}
\microtypecontext{spacing=nonfrench}

\newcommand{\ihat}{\,\hat{\imath}}
\newcommand{\jhat}{\,\hat{\jmath}}
\newcommand{\khat}{\,\hat{k}}
\newcommand{\thetahat}{\,\hat{\theta}}
\newcommand{\rhat}{\,\hat{r}}

\begin{document}
\title{Sliding down over a horizontally moving semi-sphere}
\author{Roberto~A.~Lineros$^1$}
\address{Departamento de Física, Universidad Católica del Norte, Avenida Angamos 0610, Casilla 1280, Antofagasta, Chile.}
\ead{roberto.lineros@ucn.cl}

\date{\today}

\begin{abstract}
We studied the dynamics of an object sliding down on a semi-sphere with radius $R$. 
We consider the physical setup where the semi-sphere is free to move over a flat surface. For simplicity, we assume that all surfaces are friction-less.
We analyze the values for the last contact angle $\theta^\star$, corresponding to the angle when the object and the semi-sphere detach one of each other.
We consider all possible scenarios with different combination of mass values: $m_A$ and $m_B$, and the initial velocity of the sliding object $A$.
We found that the last contact angle only depends on the ratio between the masses, and it is independent of the acceleration of gravity and semi-sphere's radius.
In addition, we found that the largest possible value of $\theta^\star$ is $48.19^{\circ}$ that coincides with the case of a fixed semi-sphere.
On the opposite case, the minimum value of $\theta^\star$ is $0^\circ$ and it occurs then the object on the semi-sphere is extremely heavy, occurring the detachment as soon as the sliding body touches the semi-sphere.
In addition, we found that if the initial kinetic energy of the sliding object $A$ is half the value of the potential energy with respect to the floor. The object detaches at the top of the semi-sphere.
\end{abstract}

\submitto{\EJP}
\maketitle
\section{Introduction}

In courses of Newtonian mechanics for engineers and physics students at undergraduate level, the concepts behind Newton's laws are key for understanding the kinematics of objects under the effects of forces. Indeed, in courses focuses on engineering, there is preference to solve problem using only Newtonian mechanics instead other possible approaches like Lagrangian or Hamiltonian mechanics. 
No matter the approach, it is troublesome for the students to understand the interplay between objects in contact due to the presence of reaction forces in cases where the contact surfaces are not flat. At the undergraduate course level, the focus on the study of vectorial mechanics is crucial for dealing with challenging problems like ones involving many bodies and the use of different vectorial basis at the same time, like cartesian, cylindrical, and spherical basis.\\

The problem of an object sliding down on a circular path is an academic example to teach such concepts~\cite{Beer, hibbeler2017, riley1996}. Similarly, the problem of an object descending on a flat surface with a slope is another common example oriented to teach relative motion in terms of relative position, velocity, and acceleration. However, sometimes there are variations on this class of problems by allowing the inclined plane to be affected by the reaction force between the object and the inclined plane (see problem 15-98 of \cite{riley1996}). In the literature, similar problems have been addressed with different approaches like the use of lagrangian mechanics~\cite{balart:2019}, including friction~\cite{prior:2007,gonzalez:2017,delpino:2018}, or the experimental demonstration~\cite{vermillion:1988}. \\

In this manuscript, we consider the situation of a moving semi-sphere, which is initially at rest, where its movement is due to the reaction force of an object sliding down on top of it. \\

The manuscript is organized as follows: In section~\ref{sec:fixed}, we present the solution for the standard case, where the semi-sphere remains still, in addition, we use the result as benchmark for the moving setup. In section~\ref{sec:mov}, We present the solution and analysis for the moving semi-sphere. Finally, section~\ref{sec:concl} are the conclusions.

%%%%%%%%%%%%%%%%%%%%%%%%%%%%%%%%%%%%%%%%%%%%%%%%%%%%%%%%%%%%%%%%%%%%%%%%%%%%%%%%
\section{System with fixed semi-sphere. \label{sec:fixed}}

\begin{figure}[t]
    \centering
    \includegraphics[width=1\textwidth]{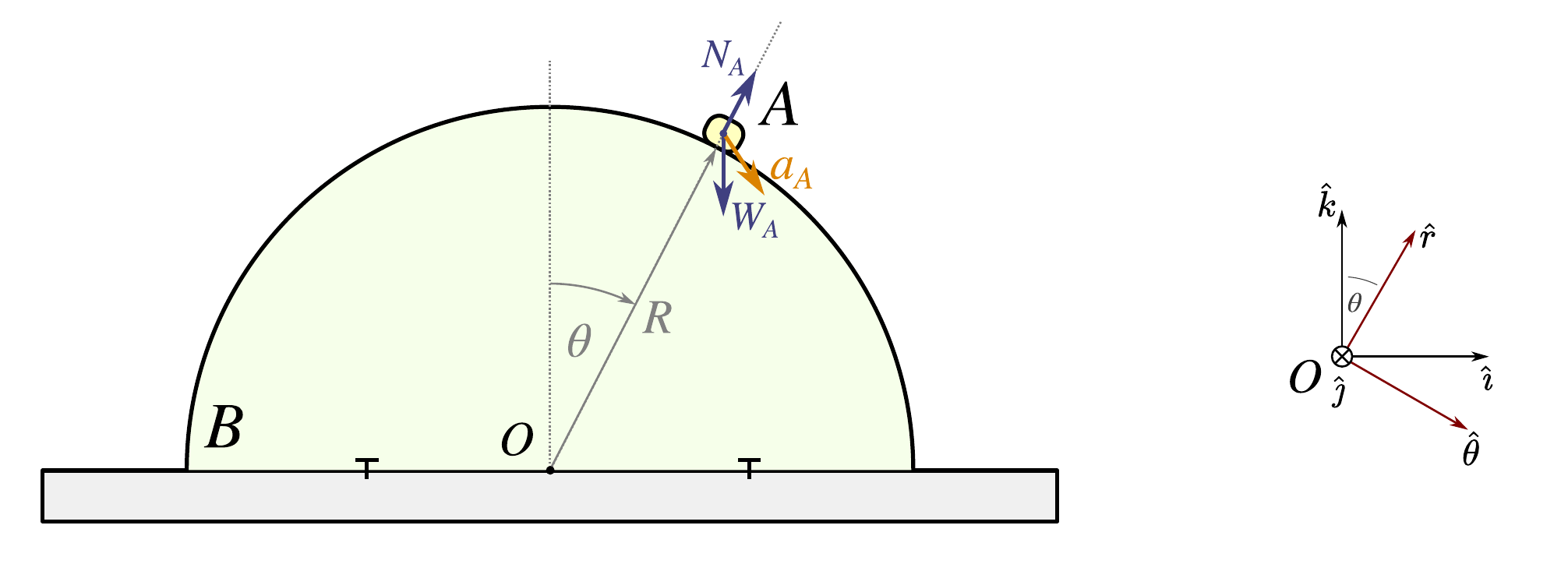}
    \caption{(Left) Physical situation on the object $A$ sliding down over a fixed semi-sphere $B$. The semi-sphere $B$ does not present friction. Forces acting on $A$ are shown in blue. (Right) Vector basis description.}
    \label{fig:fixed}
\end{figure}

At a first stage, we consider the situation when semi-sphere $B$ remains still (see figure~\ref{fig:fixed}). This is a known problem taught in courses of Newtonian Mechanics at the university level. The equations of motion for the object $A$ are constructed using the second Newton's Law and correspond to:  
\begin{eqnarray}
    \sum \vec{F}_A =  \vec{N}_A + \vec{W}_A = m_A \vec{a}_{A} \, ,
\end{eqnarray}
where $\vec{N}_A$ is the reaction force between objects and $\vec{W}_A = -m_A g \khat$ is the weight with $g$ the gravity's acceleration. While the object $A$ is in contact with the semi-sphere $B$, it moves along the surface following a circular path. The acceleration $\vec{a}_A$ is then obtained by
\begin{equation}
    \vec{a}_A = \vec{\alpha} \times \vec{r} + \vec{\omega} \times \left( \vec{\omega} \times \vec{r} \right) \, , 
\end{equation}
where $\vec{\alpha}$ and $\vec{\omega}$ are the vectors of angular acceleration and angular velocity, respectively.
For simplicity, the movement of $A$ is 2-dimensional and occurs along the plane defined by the unit vector $\ihat$ and $\khat$. Therefore, we will use a vectorial cylindrical basis defined by:
\begin{eqnarray}
    \rhat &=& \cos\theta \khat + \sin\theta \ihat \, ,\\
    \thetahat &=& -\sin\theta \khat + \cos\theta \ihat \, , 
\end{eqnarray}
where $\theta$ is the angle between $\khat$ and $\rhat$. Notice that the 3-dimensional vectorial basis maintains the right-hand rule convention, such as $\khat \times \ihat = \rhat \times \thetahat = \jhat$. In addition, this basis is different to the standard cylindrical basis, so the reader might be cautioned about this fact.

The acceleration $\vec{a}_A$ in the cylindrical basis corresponds to
\begin{equation}
    \vec{a}_A = \alpha R \thetahat - \omega^2 R \rhat \, , 
\end{equation}
when $\vec{\alpha} = \alpha \jhat$ and $\vec{\omega} = \omega \jhat$.

The equations of motions are in terms of the $\rhat$ and $\thetahat$ components:
\begin{eqnarray}
    N_A - m_A g \cos\theta & = & - m_A \omega^2 R \, , \\
    m_A g \sin\theta & = & m_A \alpha R \, .
\end{eqnarray}
For the setup in figure \ref{fig:fixed}, the angular velocity and acceleration are related to the angle $\theta$ by: 
\begin{eqnarray}
    \omega &=& \dot{\theta}\, ,\\
    \alpha &=& \frac{d \dot{\theta}}{d \theta} \dot{\theta}\, .
\end{eqnarray}
Using the latter expressions, the equations of motion are reduced to a couple of differential equations:
\begin{eqnarray}
    \frac{g}{R} \cos\theta - \frac{N_A}{m_A R} &=& \dot{\theta}^2  \, , \\
    \frac{g}{R} \sin\theta &=& \frac{d \dot{\theta}}{d \theta} \dot{\theta} \, .
\end{eqnarray}
These equations are further simplified via the substitution: $f(\theta) = \dot{\theta}^2$, $f'(\theta) = df/d\theta = 2 \ddot{\theta}$, and $\kappa = g/R$; obtaining:
\begin{eqnarray}
    \label{eq:de1} \kappa \cos\theta - \frac{N_A}{m_A R}& =& f(\theta)  \, , \\
    \label{eq:de2} 2 \kappa \sin\theta & =& f'(\theta) \,  .
\end{eqnarray}
Since the function $f(\theta)$ corresponds to the square of the angular velocity, the kinetic energy of $A$ corresponds to
\begin{eqnarray}
    T = \frac{1}{2} m_A R^2 f(\theta) \, ,
\end{eqnarray}
providing that the total mechanical energy is:
\begin{eqnarray}
    E = \frac{1}{2} m_A R^2 f(\theta) + m_A g R \cos\theta \, .
\end{eqnarray}

Depending on the program of content in a vectorial mechanics course, the concept of mechanical energy may not be seen. However, besides the resolution using equation of motion, this problem can be also solve using in addition conservation of energy.

\subsection{Solving the equations of motion}

These equations are simply solved by integrating over the $\theta$ angle in equation~\ref{eq:de2} and after by replacing in equation ~\ref{eq:de1}. When considering the initial conditions: 
\begin{eqnarray}
    \theta(t=0) &=& \theta_0 = 0\, , \\
    \dot{\theta}^2(t=0) &=& f(\theta_0) =  2 \kappa \epsilon \, ,
\end{eqnarray} 
then the reaction force and the angular velocity squared are:
\begin{eqnarray}
    N_A & = m_A g \left( 3 \cos\theta - 2(1 + \epsilon) \right)  \, , \\
    f(\theta) & = \dot\theta^2 = 2 \kappa \left( 1 + \epsilon- \cos\theta \right) \, .
\end{eqnarray}
We introduce the parameter $\epsilon$ to further simplify the solutions, however, it is related with the initial kinetic energy:
\begin{eqnarray}
    \label{eq:energyeps}
    \epsilon = \frac{T_0}{m_A g R} \, ,
\end{eqnarray}
and it can be interpreted as the ration between the initial kinetic energy and the initial potential energy.
It is important to remark that the initial position, $\theta(t=0) = 0$, is an unstable equilibrium point. In order to break the symmetrical evolution of sliding down to any side of the semi-sphere, it is important to indicate an initial direction of movement by means of the value of $\epsilon$.

The equations of motion are valid only for the regimen when $N_A \ge 0$ and describe the object $A$ moving over the semi-sphere. The case $N_A = 0$ sets the last contact angle $\theta^\star$ that in this case corresponds to:
\begin{equation}
    \label{eq:anglefixed}
    \cos\theta^\star = \frac{2}{3} \left( 1+\epsilon\right)
\end{equation}

This general solution allows us to set a maximum value of $\epsilon$ for which the object $A$ detaches at the beginning of the movement ($\cos\theta^\star = 1$):
\begin{equation}
     \epsilon_{\rm max} = \frac{1}{2} \, ,
\end{equation}
which sets the maximum angular velocity to be:
\begin{equation}
    \dot{\theta}(t=0) = \sqrt{\kappa}  \, .
\end{equation}
This sets the maximum initial kinetic energy to be:
\begin{eqnarray}
    T_0^{\rm max} = \frac{1}{2} m_A g R \, ,
\end{eqnarray}
which corresponds to half of the initial potential energy.\\

In the limit, $\epsilon \rightarrow 0$, which means that movement starts at rest, the last contact angle becomes independent of $R$ and $g$, and gives the value:

\begin{equation}
    \cos\theta^\star = \frac{2}{3}  \, \rightarrow \, \theta^\star \simeq 48.19^\circ \, .
\end{equation}

This value is a known result in the frictionless scenario, and the problem corresponds to an exercise in books~\cite{Beer, riley1996, hibbeler2017}.
A more complete case, that includes friction, can be seen in \cite{prior:2007}. This result represents a benchmark value to compare with the following case of a moving semi-sphere. 

%%%%%%%%%%%%%%%%%%%%%%%%%%%%%%%%%%%%%%%%%%
\section{System with a moving semi-sphere \label{sec:mov}}

\begin{figure}[t]
    \centering
    \includegraphics[width=1\textwidth]{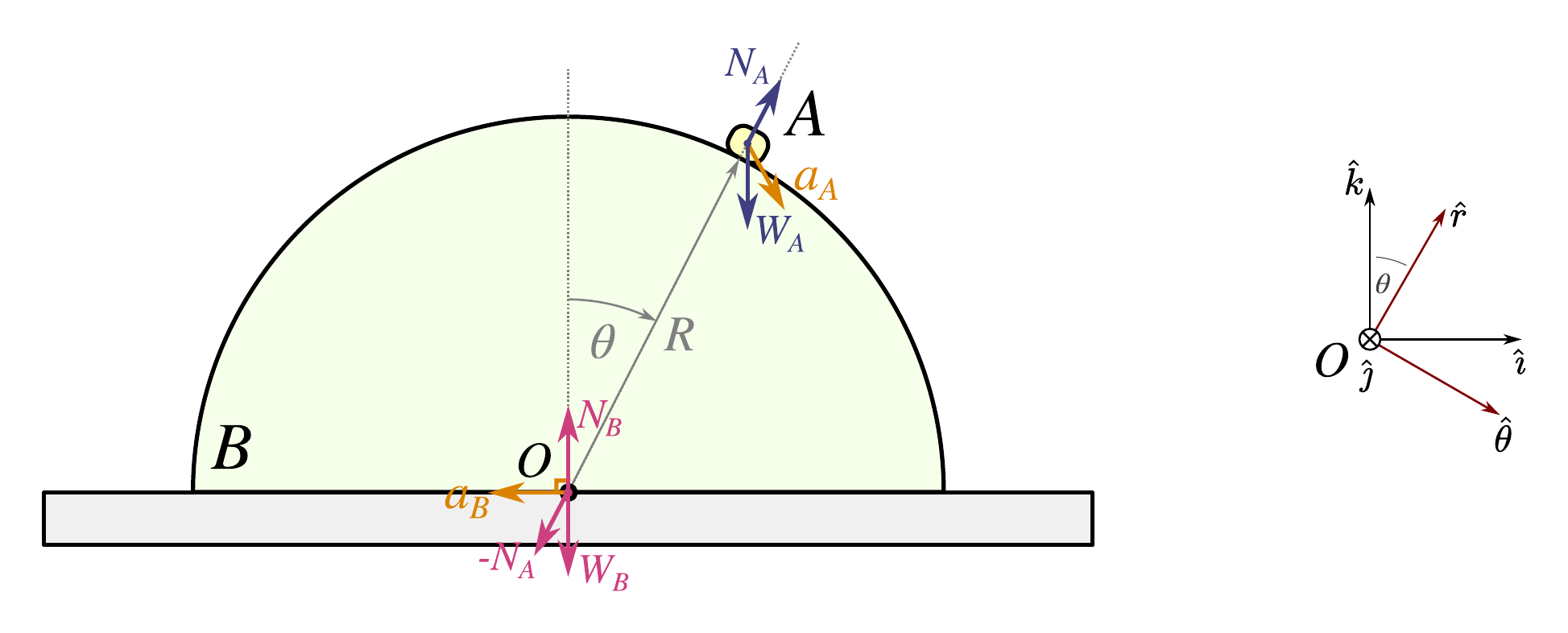}
    \caption{(Left) Physical situation on the object $A$ sliding down over a moving semi-sphere $B$. The semi-sphere $B$ does not present friction with both: object $A$, and the flat surface. Forces acting on $A$ are shown in blue. Forces acting on $B$ are shown in dark pink. Accelerations are displayed in orange. (Right) Unit vector basis description.}
    \label{fig:mov}
\end{figure}

In this part, we allow the semi-sphere to freely move over a friction-less surface. Therefore, the aim is to calculate the value of the last contact angle $\theta^\star$ under these new conditions. Figure~\ref{fig:mov} shows the physical setup.

\subsection{Equations of motions}

Similarly to the previously presented case, the object $A$ is affected by its weight and the reaction force with the semi-sphere. Therefore, the equation of motion for $A$ is:
\begin{equation}
    \sum \vec{F}_A = \vec{N}_A + \vec{W}_A = m_A \vec{a}_A \, .
\end{equation}
On the other hand, the semi-sphere $B$ is now free to move and its equation of motion is given by:
\begin{equation}
    \sum \vec{F}_B = \vec{N}_B  - \vec{N}_A + \vec{W}_B = m_B \vec{a}_B \, ,
\end{equation}
where $\vec{N}_B$ is the reaction force between the floor and the semi-sphere and $\vec{W}_B = -m_B g \khat$ is its weight. \\

The movement of the semi-sphere $B$ is only horizontal, therefore, the acceleration corresponds to $\vec{a}_B = a_{Bx} \ihat$ as well as its velocity $\vec{v}_B = v_{Bx} \ihat$. It is also important to remark that $\vec{a}_B = \vec{a}_O$ because the semi-sphere $B$ also non-rotating.\\

On the other hand, the acceleration of the object $A$ with respect to the floor i.e. the rest frame is computed through the acceleration of $B$ and the relative acceleration between objects:
\begin{eqnarray}
    \vec{a}_A &=&  \vec{a}_{B} + \vec{a}_{A/B} \, ,\\
    \vec{a}_A &=& \vec{a}_{B} + \vec{\alpha} \times \vec{r} + \vec{\omega} \times \left( \vec{\omega} \times \vec{r} \right) \, .
\end{eqnarray}
While the object $A$ is in contact with the semi-sphere $B$, the relative movement of $A$ with respect to $B$ is a circular motion. Therefore, we can write the acceleration of $A$ in term of the cylindrical components:
\begin{eqnarray}
    \vec{a}_A  &= a_{Bx} \ihat + \alpha R \thetahat - \omega^2 R \rhat \, .
\end{eqnarray}

Up to this point, the acceleration of $A$ in the cylindrical basis is:
\begin{equation}
    \vec{a}_A = \left(a_{Bx} \sin\theta - \omega^2 R\right) \rhat + \left(a_{Bx} \cos\theta + \alpha R \right) \thetahat \, ,
\end{equation}
which allow us to get the full set of equations of motion. 2 equations for the object $A$:
\begin{eqnarray}
    N_A - m_A g \cos\theta & =& m_A \left( a_{Bx} \sin\theta - \omega^2 R \right) \, , \\
    m_A g \sin\theta &=& m_A \left(a_{Bx} \cos\theta + \alpha R \right) \, , 
\end{eqnarray}
 and 2 more for the semi-sphere $B$:
\begin{eqnarray}
    -N_A \sin\theta &=& m_B a_{Bx} \, , \\
    N_B - N_A \cos\theta - m_B g &=& 0 \, .
\end{eqnarray}

After inspection, the last 2 equations lead to:
\begin{eqnarray}
    a_{Bx} &=& - \frac{N_A \sin\theta}{m_B} \, ,\\
    N_B &=& m_B g + N_A \cos\theta \, ,
\end{eqnarray}
and these can be used to reduce the full set of equations of motion to only two equations:
\begin{eqnarray} 
    \label{eq:combeom1}
    \frac{g}{R} \cos\theta - \frac{N_A}{m_A R} \left( 1 + \frac{m_A \sin^2\theta}{m_B}\right) &=& f(\theta) \, ,\\
    \label{eq:combeom2}
    \frac{2g}{R} \sin\theta + \frac{2 N_A}{m_B R} \sin\theta \cos\theta &=& f'(\theta) \, ,
\end{eqnarray}
where $f(\theta) = \dot{\theta}^2$ and $f'(\theta) = 2 \ddot{\theta}$. Notice, we use the same relations for the angular acceleration and velocity with respect to the angle $\theta$: $\alpha = \ddot{\theta}$, and $\omega = \dot{\theta}$.\\

The function $f(\theta)$ is related to the kinetic energy of $A$ only if the semi-sphere $B$ is at rest. In a general case, the kinetic energy of $A$ depends on the velocity of the semi-sphere and the relative velocity between objects:
\begin{eqnarray}
    T_A = \frac{1}{2} m_A v_A^2 = \frac{1}{2} m_A \left(\vec{v}_B + \vec{v}_{A / B} \right)^2 = \frac{1}{2} m_A \left(\vec{v}_B + \vec{\omega} \times \vec{r}\right)^2 \, ,
\end{eqnarray}
where in terms of $f(\theta)$, it corresponds to:
\begin{eqnarray}
    T_A = \frac{1}{2} m_A \left( v_B^2 + 2 v_B \cos\theta R \sqrt{f(\theta)} + R^2 f(\theta) \right) \, ,
\end{eqnarray}
and with it, we can obtain the total mechanical energy of the system:
\begin{eqnarray}
    E = \frac{m_B}{2}  v_B^2 + \frac{m_A}{2}\left( v_B^2 + 2 v_B \cos\theta R \sqrt{f(\theta)} + R^2 f(\theta) \right) + m_A g R \cos\theta \, .
\end{eqnarray}

%%%%%%%%%%%%%%%%%%%%%%%%%%%%%%%%%%%%%%
\subsection{Solving the equations of motion}

The solution of the system of equation cannot be performed as in the fixed semi-sphere case because the reaction force $N_A$ is present in both equations, and it depends on the angle. Nevertheless, the reaction force $N_A$ can be isolated from the equation \ref{eq:combeom1} and be written in terms of the dynamical variables:
\begin{equation}
    N_A(\theta) = m_A R \, \frac{ \kappa  \cos\theta - f(\theta )}{1+ \beta  \sin^2\theta} \, ,
\end{equation}
where $\beta = m_A/m_B$ is the ratio between the masses and $\kappa = g/R$ is the ratio between the acceleration of gravity and the radius of the semi-sphere. Here, the reaction force depends on the square of the angular velocity encoded in $f(\theta)$.

After removing the explicit dependence of $N_A$ in the equation \ref{eq:combeom2}, we get the differential equation for $f(\theta)$:
\begin{eqnarray}
    \frac{1+ \beta  \sin^2\theta}{2 \sin\theta} \, f'(\theta)  + \beta \cos\theta \, f(\theta) - \kappa (1 +\beta) = 0 \, .
\end{eqnarray}
This differential equation is analytically solvable and has the solution:
\begin{eqnarray}
    \label{eq:solde}
    f(\theta) = \frac{2 \kappa (1 + \beta) (1 - \cos\theta) + \epsilon}{1+ \beta  \sin^2\theta} \, ,
\end{eqnarray}
when the initial conditions: $\theta(t=0) = 0$ and $f(\theta_0) =  2 \kappa \epsilon$ are included. Here, the parameter $\epsilon$ is related to the kinetic energy of the object $A$ in the same way as in the previous section (equation \ref{eq:energyeps}). This is because the semi-sphere $B$ is initially at rest.\\

Let's remark that this solution is valid for, $N_A(\theta) \ge 0$  and it holds for angles $0\le\theta\le\theta^\star\le \pi/2$ where $\theta^\star$ corresponds to the last contact angle.

%%%%%%%%%%%%%%%%%%%%%%%%%%%%%%%%%%%%%%
\subsection{Finding the last contact angle}

The angle $\theta^\star$ can be obtained by solving the following equation:
\begin{equation}
    \label{eq:NA}N_A(\theta^\star) = m_A R \, \frac{ \kappa  \cos\theta^\star - f(\theta^\star )}{1+ \beta  \sin^2\theta^\star}  = 0 \, ,
\end{equation}
where $f(\theta^\star)$ is the solution of the differential equation evaluated at $\theta^\star$ (equation \ref{eq:solde}). This leads to the following equation to solve for $\theta^\star$:
\begin{eqnarray}
    \label{eq:precubic}
    \kappa \cos\theta^\star = f(\theta^\star) = \frac{2 \kappa (1 + \beta) (1 - \cos\theta^\star) + \epsilon}{1+ \beta  \sin^2\theta^\star}\, .
\end{eqnarray}
This equation can be written as a depressed cubic equation for $\xi = \cos\theta^\star$ as follows: 
\begin{equation}
      \label{eq:dep} 
      H(\xi) = \sin^2{\left(\frac{\pi}{2}\tau\right)} \,\xi^3 - 3 \xi  + 2 + 2\epsilon \cos^2{\left(\frac{\pi}{2}\tau\right)} = 0 \, ,
\end{equation}
where the $H(\xi)$ function is just a reparametrization of equation \ref{eq:precubic} in terms of adimensional parameters.\\

\begin{figure}[t]
    \centering
    \includegraphics[width=1\textwidth]{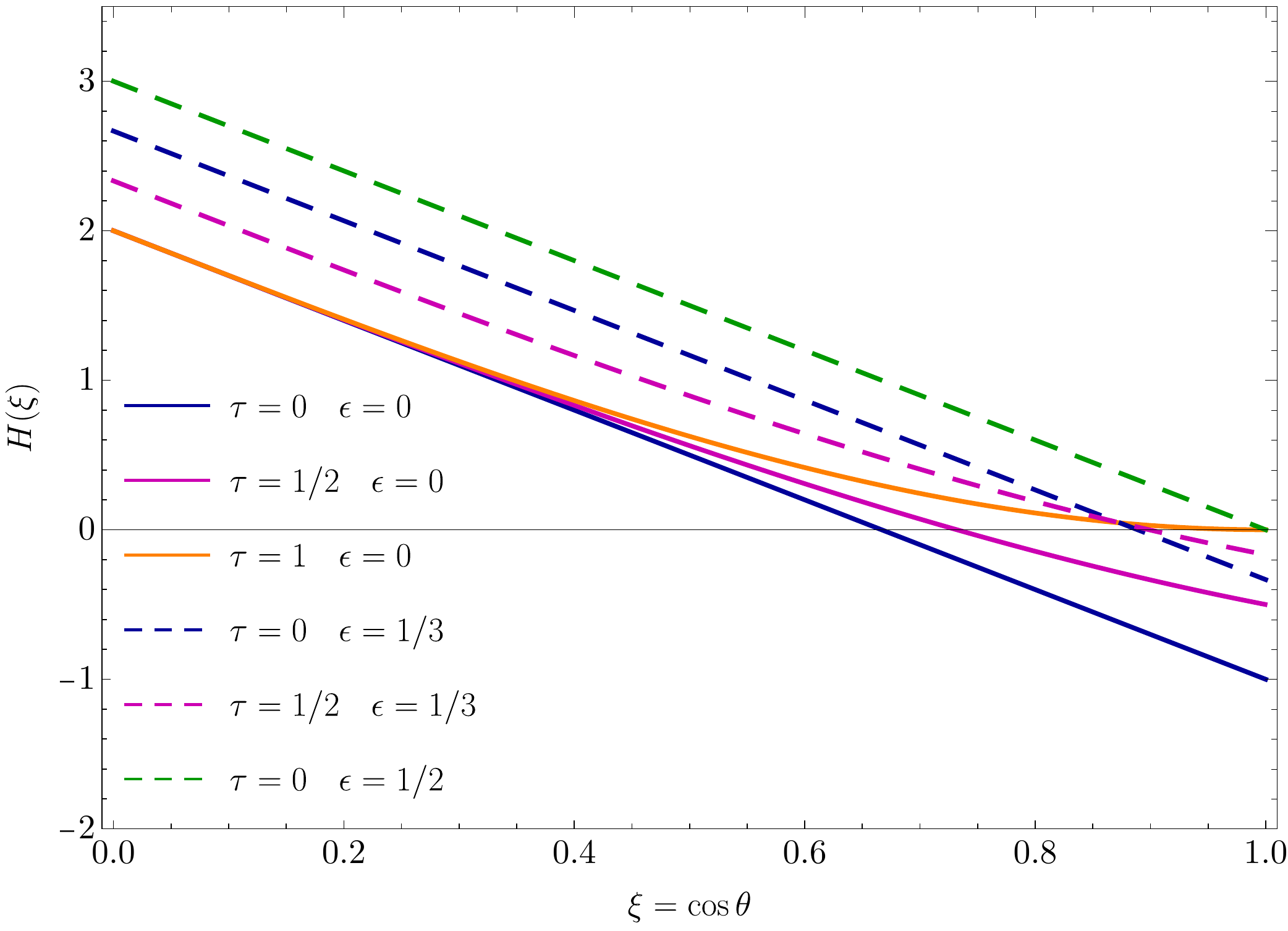}
    \caption{\label{fig:hxi} $H(\xi)$ versus $\xi$ for combination of the parameters: $\tau = 0, 1/2, 1$ and $\epsilon = 0$ (solid lines), $1/3$ (dashed lines). The green dashed line correspond to $\epsilon=1/2$ and $\tau=0$. The intersection of each curve $H(\xi)$ with the horizontal black line ($H(\xi)=0$) produces the solution where $\xi = \cos\theta^\star$.}
\end{figure}

We introduce the $\tau$-parameter ranging $0\le \tau\le 1$ that simplifies the expressions better than the mass ratio $m_A/m_B$ such as:
\begin{equation}
	\beta = \frac{m_A}{m_B} = \tan^2\left(\frac{\pi}{2} \tau \right) \, .
\end{equation}
At the extremes values of $\tau$, we obtain that $\tau \rightarrow 0$ corresponds to $\beta \rightarrow 0$ meaning the case of a heavy semi-sphere which is equivalent to the fixed semi-sphere case. Similarly, when $\tau \rightarrow 1$ corresponds to $\beta \rightarrow \infty$ meaning the scenario where object $A$ is extreamly heavy with respect to the semi-sphere. In figure~\ref{fig:hxi}, we present the function $H(\xi)$ for various values of $\tau$. \\ 

The analysis of the equation~\ref{eq:dep} reveals the maximum value that $\epsilon$ can take. To find this, we evaluate the function $H(\xi)$ in $\xi=1$ ($\theta^\star = 0$):
\begin{equation}
    H(1) = (2 \epsilon_{\rm max} - 1) \cos^2{\left(\frac{\pi}{2} \tau\right)} = 0\, ,
\end{equation}
obtaining that for the range $0\leq \tau < 1$, the maximum value is:
\begin{equation}
    \epsilon_{\rm max} = \frac{1}{2}\, ,
\end{equation}
which is same the same limit obtained in the fixed semi-sphere case. 
This means that the kinetic energy of $A$ can be at most half of the initial potential energy in order to get the object $A$ sliding down the semi-sphere $B$. This is regardless of the mass ratio $\beta$. \\

As stated before, the limit of fixed semi-sphere is reached when $\tau \rightarrow 0$ and it corresponds to the mass limit: $m_A \rightarrow 0$ or $m_B \rightarrow \infty$. In this limit, equation \ref{eq:dep} corresponds to:
\begin{equation}
    - 3 \xi  + 2 + 2 \epsilon = 0 \, ,
\end{equation}
with solution :
\begin{eqnarray}
    \xi = \cos\theta^\star = \frac{2}{3} (1+\epsilon) \, ,
\end{eqnarray}
corresponding exactly to equation \ref{eq:anglefixed}.
This case agrees with the solution for the fixed semi-sphere described in section~\ref{sec:fixed}, and it indicates the limit $\beta$ is equivalent to restrict the movement of $B$ to be fixed in a point.\\

The other limit, $\tau \rightarrow 1$, gives the equation:
\begin{equation}
    \xi^3 - 3 \xi + 2 = (\xi + 2) (\xi - 1)^2 = 0\, , 
\end{equation}
with 3 real solutions: $\xi_1 = 1$, $\xi_2 = 1$, and $\xi_3 = -2$. The solution with physical meaning are $\xi_{1,2}$, both correspond to an angle $\theta^\star = 0$. This means when $m_A \rightarrow \infty$ the angle of last contact between $A$ and $B$ is at the top of the semi-sphere and happens at the beginning of the movement. In this case, the semi-sphere moves fast enough to not be in contact with the object $A$.\\

For the intermediate cases, $0<\tau<1$ and $\epsilon=0$, the solutions for equation~\ref{eq:dep} corresponds to:
\begin{eqnarray}
    \xi_1 &=& \frac{2}{1 + 2 \cos{\left( \frac{\pi}{3} \tau\right) } }\, , \\
    \xi_2 &=& \frac{\sqrt{3} \cos{\left(\frac{\pi}{6} \tau \right)} - \sin{\left(\frac{\pi}{6} \tau \right)}}{\sin{\left(\frac{\pi}{2} \tau \right)}}\, , \\
    \xi_3 &=& - \frac{\sqrt{3} \cos{\left(\frac{\pi}{6} \tau \right)} + \sin{\left(\frac{\pi}{6} \tau \right)}}{\sin{\left(\frac{\pi}{2} \tau \right)}} \, , \\
\end{eqnarray}
which are obtained by analytically solving the depressed cubic equation. However, in order to get the real values the solutions of equation~\ref{eq:dep} need to be rephased by $e^{i \pi/3}$ when the equation is solved via the Vieta's substitution~\cite{354013610X}. From the 3 roots, $\xi_1$ has a physical meaning: $\xi_1 = \cos\theta^\star$ while the other 2 roots give values outside the physical range $0 \le \xi \le 1$. \\

The approach to analytically solve $H(\xi) = 0$ might be hard to perform by students. An alternative approach and much easier might be by using a numerical code, for example, written in python (See appendix \ref{sec:python}).\\

\begin{figure}[t]
    \centering
    \includegraphics[width=1\textwidth]{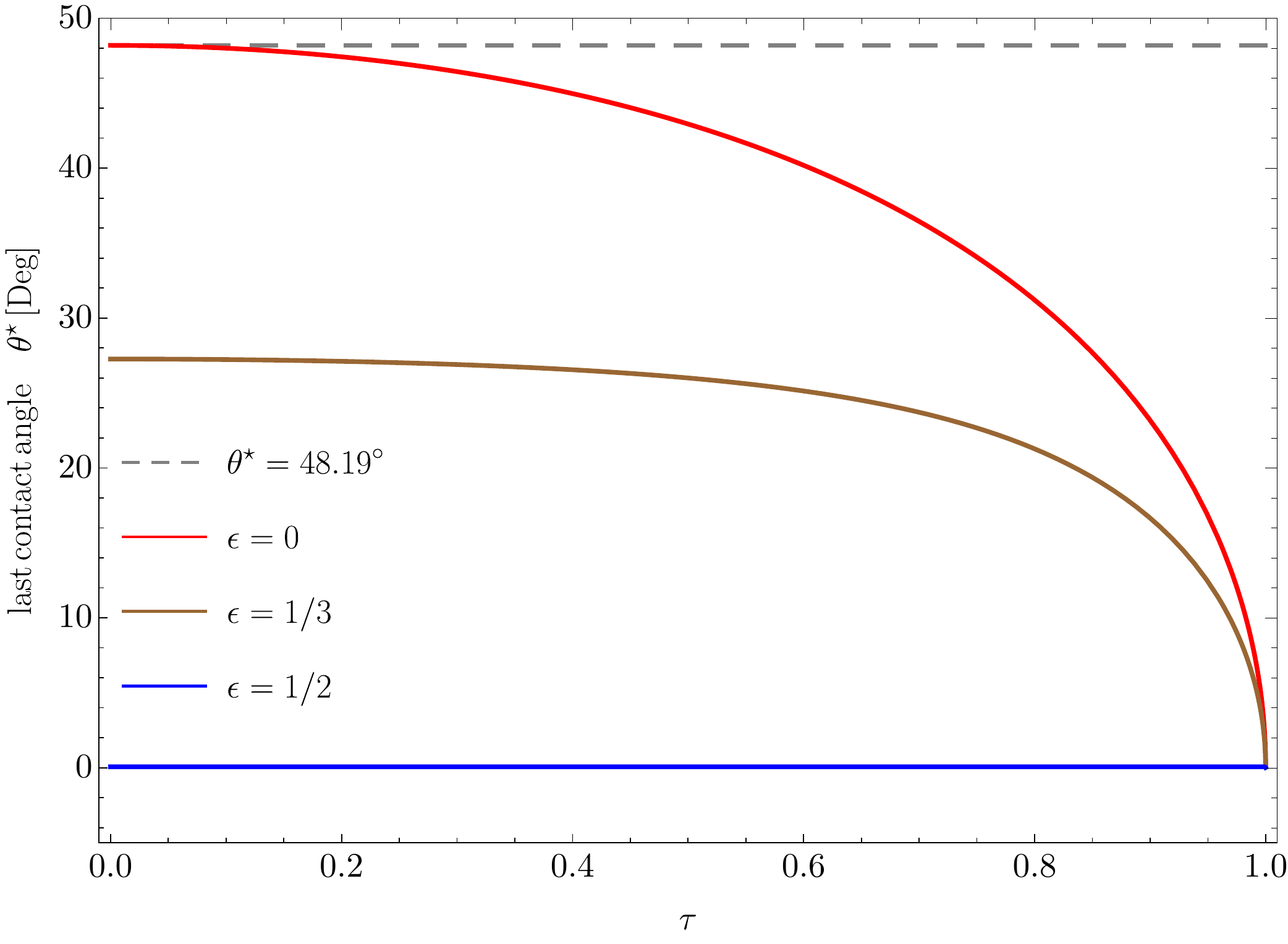}
    \caption{\label{fig:thetastar} Last contact angle $\theta^\star$ versus $\tau$. Red line shown the functional dependence in $\tau$ for $\epsilon=0$, the brown line corresponds to the case when $\epsilon=1/3$, and blue line for $\epsilon=1/2$. The gray dashed line indicate $\theta^\star=48.19^\circ$ which corresponds to the solution of the fixed-sphere case. }
\end{figure}

The dependence of the last contact angle $\theta^\star$ in terms of $\tau$ and $\epsilon$ is shown in figure~\ref{fig:thetastar}. We observe that the solution of the cubic equation include the extreme limits, like the fixed semi-sphere ($\tau=0$) for different values of $\epsilon$.\\

In addition, the value of $\theta^\star$ indicates that the largest possible value for any configuration of masses corresponds to the fixed semi-sphere case and the lowest corresponds when $m_A \rightarrow \infty$ or $m_B \rightarrow 0$ producing an extreme situation where the semi-sphere $B$ and the object $A$ get immediately detached upon the first contact.

Moreover, the dependence of $\theta^\star$ in terms of the value of $\epsilon$ gives a interesting situation that implies that the object $A$ can only remain in contact with the semi-sphere if its initial kinetic energy is, at most, half of the initial potential energy.
This can be appreciated in figure~\ref{fig:thetastar} where the blue line gives the value $\theta^\star = 0$ for all possible values of $\tau$.

%%%%%%%%%%%%%%%%%%%%%%%%%%%%%%%%%%%%%%%
\section{Conclusions \label{sec:concl}}

We present the analytical solution of finding the last contact angle for the problem of an object sliding down on a semi-sphere of radius $R$ where the semi-sphere is on a friction-less surface. The approach used to solve the problem is in terms of Newtonian mechanics and concepts of vectorial mechanics, which are topics familiar to undergraduate students of engineering and bachelor in physics. The key effect to consider is the reaction forces between the object and the semi-sphere. This force provokes the semi-sphere to move horizontally and the object $A$ to descend, keeping contact with the semi-sphere. If the velocity of the object $A$ is larger enough that the reaction force between the object and the surface is null, then both objects detach one of each other. The angle in which that occurs is the last contact angle, and it was calculated analytically.\\

We found that the case of a fixed semi-sphere (equivalent to $m_B \gg m_A$) gives the maximum possible last contact angle among any physical configuration of masses. In addition, we found that this angle depends only on the ratio of the masses $m_A$ and $m_B$ and it is independent of the value of the acceleration of gravity or the radius of the semi-sphere.\\

In addition, we include the effect produced by the initial velocity that the sliding object might have. This effect is parameterizes in terms of the quantity, $\epsilon$, which is the ratio between the initial kinetic energy and the potential energy. We found also that if the kinetic energy is larger than 1/2 of the initial potential energy, then the sliding object detaches at the very beginning of the movement.\\

The problem discussed in this manuscript present a general physical scenario that might be worth to be used as an example for enthusiast students in courses of Newtonian mechanics at the undergraduate or graduate level. In addition, to verify experimentally might also be interesting.

\ack
We thank Nicolas Rojas, Fernando Guzmán, Julio Yañez, and Eduardo Peinado for useful discussions and comments. 
We also thank the comments and suggestions from the anonymous referees.
RL is supported by Universidad Católica del Norte through the Publication Incentive program No. CPIP20180343 and CPIP20200063.

\appendix
\section{Python code}
\label{sec:python}

Here, we present an example of a code in python for solving equation \ref{eq:dep} and return the value of the last contact angle in degrees.

\lstset{frame=tb,
  language=Python,
  aboveskip=3mm,
  belowskip=3mm,
  showstringspaces=false,
  columns=flexible,
  basicstyle={\small\ttfamily},
  numbers=none,
  numberstyle=\tiny\color{gray},
  keywordstyle=\color{blue},
  commentstyle=\color{gray},
  stringstyle=\color{red},
  breaklines=true,
  breakatwhitespace=true,
  tabsize=1
}

\begin{lstlisting}
import numpy as np
import scipy as sp 
import matplotlib.pyplot as plt 
import scipy.optimize as spop

def H(xi,tau,eps):
    PI = np.pi #value of pi
    S2 = (np.sin(PI*tau/2))**2 #sin^2
    C2 = (np.cos(PI*tau/2))**2 #cos^2
    out = S2*xi**3 - 3*xi + 2 + 2*eps*C2 # function H(xi)
    return out

# initial values for tau and epsilon
tau = 0.0 #tau
eps = 0.5 #epsilon
root = spop.fsolve(H, 0.5, args=(tau,eps))

print("The last contact angle for eps =", eps, "and tau =",tau, "is:", np.arccos(root[0])*180/np.pi, "degrees")
\end{lstlisting}

%%%%%%%%%%%%%%%%%%%%%%%%%%%%%%%%
%%%%%%%% REFERENCES
%%%%%%%%%%%%%%%%%%%%%%%%%%%%%%%%
%%%%%%%% Journals %%%%%%%%%%%%%%
\def\apj{Astrophys.~J.}                       % Astrophysical Journal
\def\apjl{Astrophys.~J.~Lett.}                % Astrophysical Journal, Letters
\def\apjs{Astrophys.~J.~Suppl.~Ser.}          % Astrophysical Journal, Supplement
\def\aap{Astron.~\&~Astrophys.}               % Astronomy and Astrophysics
\def\aj{Astron.~J.}                           %
\def\araa{Ann.~Rev.~Astron.~Astrophys.}       %
\def\mnras{Mon.~Not.~R.~Astron.~Soc.}         %
\def\physrep{Phys.~Rept.}                     %
\def\jcap{J.~Cosmology~Astropart.~Phys.}      % Journal of Cosmology and Astroparticle Physics
\def\jhep{J.~High~Ener.~Phys.}                % Journal of High Energy Physics
\def\prl{Phys.~Rev.~Lett.}                    % Physical Review Letters
\def\prd{Phys.~Rev.~D}                        % Physical Review D
\def\nphysa{Nucl.~Phys.~A}                    % Nuclear Physics A

\section*{References}
\bibliographystyle{iopart-num}
\bibliography{references.bib}

\end{document}